# Photovoltaic effect in an electrically tunable van der Waals heterojunction

Marco M. Furchi[1], Andreas Pospischil[1], Florian Libisch[2], Joachim Burgdörfer[2], and Thomas Mueller[1,*]

[1]*Vienna University of Technology, Institute of Photonics,*
*Gußhausstraße 27-29, 1040 Vienna, Austria*
[2]*Vienna University of Technology, Institute for Theoretical Physics,*
*Wiedner Hauptstraße 8-10, 1040 Vienna, Austria*

**ABSTRACT: Semiconductor heterostructures form the cornerstone of many electronic and optoelectronic devices and are traditionally fabricated using epitaxial growth techniques. More recently, heterostructures have also been obtained by vertical stacking of two-dimensional crystals, such as graphene and related two-dimensional materials. These layered designer materials are held together by van der Waals forces and contain atomically sharp interfaces. Here, we report on a type-II van der Waals heterojunction made of molybdenum disulfide and tungsten diselenide monolayers. The junction is electrically tunable and under appropriate gate bias, an atomically thin diode is realized. Upon optical illumination, charge transfer occurs across the planar interface and the device exhibits a photovoltaic effect. Advances in large-scale production of two-dimensional crystals could thus lead to a new photovoltaic solar technology.**

The discovery of the photovoltaic effect in a silicon p-n junction by Chapin, Fuller, and Pearson[1] heralded the era of photovoltaics. Although today's technology is still mainly based on silicon, a multitude of other materials have been investigated since then, with perovskite[2] being one of the most recent additions. Lately, a new class of materials has emerged – two-dimensional (2D) atomic crystals[3, 4]. These can be produced with large area dimensions at low costs and on flexible substrates[5–7], making them potentially attractive candidates for solar energy conversion.

Most solar cell designs comprise an interface (junction) between two adjoining materials, at which the separation of the photo-generated charge carriers occurs. Schottky and p-n junctions have been realized in atomically thin transition metal dichalcogenides (TMDCs) and the capability of 2D crystals for photovoltaics has been demonstrated[8–11]. However, the lateral arrangement of these devices does not allow for easy scalability, for which a vertical geometry would be desirable. Such vertical heterostructures can be obtained by stacking of 2D crystals in a layered configuration on top of each other[12]. The van der Waals interaction between adjacent layers keeps the stack together.

These so-called van der Waals heterostructures constitute a new class of designer materials that can be composed of a variety of 2D crystals that are structurally and mechanically similar but exhibit entirely different electronic and optical properties. Hexagonal boron nitride (hBN), for instance, is a wide-gap insulator and optically transparent. It has been employed as electronic barrier in tunneling devices[13, 14] and as crystalline substrate[15, 16]. Graphene is a semi-metal and is typically used as conducting electrode. Another group of 2D crystals are TMDCs, which have the formula $MX_2$, where M is a transition metal and X = S, Se, or Te. TMDCs exhibit semiconducting behavior[17–

[20] and have demonstrated considerable potential for acting as active electronic[21–25] and optoelectronic[21, 22, 26–30] components.

Photodetectors based on graphene/TMDC/graphene stacks[29, 30] have shown excellent performance with external quantum efficiency ($EQE$) of up to 55 %. However, such a configuration would not be well suited for photovoltaic energy conversion, because of the low resistance of the unipolar TMDC channel that bypasses the photocurrent source. The generated photovoltage thus varies strongly with illumination intensity, and almost vanishes at low light levels, making the total electric power too small for practical applications. By contrast, the photovoltage in a diode configuration is determined by the relative band alignments at the p-n junction and varies only weakly with illumination intensity. In this letter, we demonstrate that a van der Waals heterojunction made of two different dichalcogenides behaves as a diode and exhibits a photovoltaic effect. The device can thus be employed for solar-energy conversion.

Our devices (see Figure 1a for a schematic drawing) were fabricated on an oxidized silicon (Si) wafer with a 280 nm thick silicon dioxide ($SiO_2$) layer. They consist of molybdenum disulfide ($MoS_2$) and tungsten diselenide ($WSe_2$) monolayers that were mechanically exfoliated from bulk crystals and stacked on top of each other. In a first step, the bottom-layer flake was exfoliated directly onto the $Si/SiO_2$ substrate, identified by optical microscopy, its thickness verified by Raman (see Supporting Information, Figure S1) and photoluminescence (PL) spectroscopy, and annealed for several hours in vacuum ($\sim 5\times 10^{-6}$ mbar) at a temperature of $T = 380$ K. In a second step, the top-layer flake was exfoliated onto a stack of polymers on a sacrificial silicon wafer. The polymer stack consisted of polyacrylic acid (PAA) and polymethyl methacrylate (PMMA), with a thickness chosen such that a monolayer flake could again be identified using a microscope. The PAA was then dissolved in water and the PMMA with the monolayer flake was released from the wafer, turned upside down, and placed with micrometer-precision on top of the other flake, with the target wafer kept at $T = 380$ K (above the glass transition temperature of PMMA). Finally, the wafer was cooled down to room temperature and the PMMA was dissolved. Short circuits, caused by chunks of bulk material that are inevitably transferred with the monolayer flake, were removed using a focused ion beam (FIB). After sample fabrication the devices were annealed again for several hours at $T = 380$ K in vacuum (~$5\times10^{-6}$ mbar) to remove resist residues and other contaminants from the sample. A gate voltage $V_G$, applied to the silicon substrate, controls the electrical characteristics of the heterojunction device. We investigated samples (denoted I, II and III in the following) with both $MoS_2/WSe_2$ and $WSe_2/MoS_2$ layer orders. They all displayed qualitatively similar behavior. Figure 1b shows a microscope image of sample I. A scanning electron microscopy (SEM) image of the device is presented in Figure S2.

Before we discuss the device operation, let us recall some basic properties of the materials used. Figure 1c shows schematic energy band diagrams of $MoS_2$ (left) and $WSe_2$ (right), respectively. Both are direct semiconductors when thinned to monolayer thickness[19, 31, 32] with conduction band minima and valence band maxima at the $K$- and $K'$-points. The absorption spectra of these materials approximately match the solar spectrum, so that they are both suitable as optical absorbers in a photovoltaic cell[33]. In Figure 1d we show PL spectra obtained by focusing the light from a solid-state laser (532 nm wavelength) onto three different positions on the sample ($WSe_2$ monolayer, red line;

MoS$_2$ monolayer, blue line; MoS$_2$/WSe$_2$ heterojunction, yellow line). The individual WSe$_2$ and MoS$_2$ monolayers show strong excitonic emission at ~1.65 eV and ~1.85 eV (A-exciton; ~2.0 eV, B-exciton), respectively, in agreement with previous reports[19, 31, 32]. Also the higher quantum yield of WSe$_2$ as compared to MoS$_2$ has been reported frequently. The suppression of the MoS$_2$ A-peak with respect to the B-peak indicates strong n-doping of the MoS$_2$ crystal[34] (see also electrical measurements). The emission from the heterojunction is quenched – as we will discuss below. Given the large exciton binding energies of $E_b$ ~ 0.5 eV in WSe$_2$ and MoS$_2$ 2D materials[35–37], we estimate electronic band gaps of $E_{GW}$ ~ 2.15 eV and $E_{GM}$ ~ 2.35 eV. The electron affinity (the energy required to excite an electron from the bottom of the conduction band to vacuum) of MoS$_2$ is[21] $\chi_M$ ~ 4.2 eV. Values reported for monolayer WSe$_2$ vary, but are generally lower[23] ($\chi_W$ ~ 3.5–4.0 eV), owing to the smaller electronegativity of W and Se compared to Mo and S.

When brought into contact, an atomically-sharp van der Waals heterojunction is formed that is pristine and without broken bonds. The center part of Figure 1c shows the band diagram of the heterojunction, where we assume flat band alignment with the neighboring monolayer regions for simplicity. The exact band lineup will not only be governed by the intrinsic properties of the materials but also by extrinsic factors that vary from sample to sample and are hence difficult to predict. For instance, unintended n-doping is typically observed in natural MoS$_2$; adsorbed molecules from the ambient will give rise to additional doping. The proposed band diagram can hence only provide a qualitative picture. Because of the large energetic offset of the MoS$_2$ and WSe$_2$ states, interlayer coupling is negligible and the heterostructure bands can be considered as a superposition of the monolayer bands[38]. Consequently, the lowest energy electron states are spatially located in the MoS$_2$ layer and the highest energy hole states lie in the WSe$_2$, thus forming a type-II heterostructure, as confirmed by band structure calculations[38–40] and recent PL measurements[41].

In a first step after device fabrication we acquired the electrical characteristics, where the silicon substrate served as a back gate electrode to adjust the doping in the device. Figure 2a shows the current map as a function of gate ($V_G$) and bias ($V$) voltages, recorded under biasing conditions as shown in Figure 1a (sample I). The electrical characteristics are different from those obtained in ordinary MoS$_2$ or WSe$_2$ field-effect transistors[17, 18, 20, 23], but can readily be understood by considering an electrical series connection of two such transistors[42] (see also Figure S3). While the MoS$_2$-channel remains n-type over most of the observed gate voltage range and gets fully depleted only at $V_G$ < -71 V, the WSe$_2$-channel switches from n-type ($V_G$ > -11 V) to p-type ($V_G$ < -47 V) due to its ambipolar behavior. In the range -47 V < $V_G$ < -11 V the WSe$_2$-sheet becomes intrinsic and the current flow is suppressed. The less efficient gating of MoS$_2$ as compared to WSe$_2$ is mainly attributed to the existence of an impurity band tail underneath the conduction band edge through which the Fermi level moves slowly[18]. The electrical transport in the MoS$_2$ then occurs by carriers that are thermally activated from impurity states to the conduction band[43].

The $V$-dependence of the current is strikingly different in the two on-states. At $V_G$ > -11 V, when MoS$_2$ and WSe$_2$ are both n-type, current can flow under both $V$-polarities. In the range -71 V < $V_G$ < -47 V, however, the current-voltage ($J$–$V$) characteristic exhibits a nonlinear response with current flow only under forward ($V$ > 0) bias. $J$–$V$ traces obtained under n-n ($V_G$ = +10 V) and p-n ($V_G$ = -59 V) configurations are shown in Figure 2b as dashed blue and solid red lines, respectively (the data are also provided on a linear scale in Figure S4). While the *J–V* curve of the n-n junction shows symmetric (resistive) behavior, the p-n configuration displays diode-like current rectification, consistent with the type-II band alignment. The forward/reverse current ratio is ∼50 at |*V*| = 8 V. The electrical characteristics of our device can hence be controlled by electrostatic doping (see also Figure S5).

Better insight in the diode operation is gained by the band diagram inset in Figure 2b. Forward biasing the diode drives electrons (holes) in $MoS_2$ ($WSe_2$) to the heterojunction, where they may either (i) overcome the conduction (valence) band offset $\Delta E_C$ ($\Delta E_V$) to be injected into the neighboring layer and drift to the opposite contact electrode, or (ii) recombine so that a continuous current is maintained. At this point it also becomes clear why we have chosen Pd as a contact metal. Pd has a high-work function which allows for efficient hole injection[44] into $WSe_2$, but still serves as a reasonably good n-type contact to $MoS_2$ due to the strong Fermi level pinning in this material[45].

Figure 3a shows *J–V* curves upon illumination with white light from a halogen lamp with incident optical power $P_{opt}$ varied between 180 and 6400 W/m² (sample II; see electrical characteristics in the dark in Figure S6). The optical excitation spectrum is provided in Figure S7. In all measurements, the device is operated in the p-n regime with $V_G$ fixed at -50 V. From the *J–V* traces it is obvious that the device shows a photovoltaic response, as the curves pass through the fourth quadrant, and electrical power $P_{el}$ can be extracted (Figure 3b). The illumination dependence of the diode series resistance is attributed to photo-gating of MoS$_2$. In brief, optically excited carriers are captured in long-lived trap states and increase the MoS$_2$ conductivity[26].

Our physical picture of the photoresponse is the following: Photons are absorbed in WSe$_2$ and MoS$_2$, resulting in electron-hole pairs (excitons) in both monolayers. Relaxation of the photogenerated carriers then occurs, driven by the type-II band offsets $\Delta E_C$ and $\Delta E_V$. As the lowest energy electron and hole states are spatially separated, charge transfer occurs across the heterojunction. The relaxed carriers then diffuse laterally to the contacts, resulting in a photocurrent. Interlayer recombination may occur during diffusion, which reduces the efficiency of the solar cell (in fact, this appears to be the limiting factor for device performance). In the inset of Figure 3a, we schematically illustrate this process for a photon that is absorbed in the WSe$_2$ layer (an analog diagram can be drawn for photons absorbed in MoS$_2$ – not illustrated, for clarity).

The PL data in Figure 1d support the above interpretation. The luminescence from both layers in the overlap region is quenched (yellow line), indicating strong interlayer coupling. The PL results are inconsistent with type-I band alignment, for which PL quenching for the higher-band gap material (MoS$_2$) only would be expected. The WSe$_2$ luminescence is reduced by ~98 %, whereas the reduction in PL from MoS$_2$ is only ~65 %. This behavior is understood as a result of competition between interlayer charge transfer and (non-radiative) intralayer recombination. The lower luminescence yield of MoS$_2$ as compared to WSe$_2$ implies faster intralayer recombination in the former and therefore less efficient charge transfer.

From a molecular perspective, used to describe organic solar cells, our device can be regarded as a heterojunction cell, in which the MoS$_2$ conduction ($E_{CM}$) and WSe$_2$ valence ($E_{VW}$) bands act as LUMO and HOMO levels, respectively. However, in contrast to organic solar cells, no exciton transport by diffusion is required, because of electron-hole pair generation right at the junction. A spatially resolved photocurrent map (see Figure S8) was recorded using a tightly focused laser spot (532 nm wavelength) to verify that the photoresponse indeed stems from the laterally extended heterojunction and is not produced at one of the metal-semiconductor contacts[8].

In Figures 3c and d we present short-circuit current $J_{SC}$ and open-circuit voltage $V_{OC}$ as a function of illumination intensity (extracted from Figure 3a). $J_{SC}$ scales linearly with $P_{opt}$ and shows no indication of saturation over the measured range of optical powers, which extends to intensities ~5 times higher than the terrestrial solar intensity. We calculate the external quantum efficiency $EQE$, which is the ratio of the number of collected charge carriers to the number of incident photons, from $J_{SC} = EQE \times \lambda q/(hc) \times P_{opt}$ ($\lambda \sim 590$ nm is the wavelength, and $q$, $h$, and $c$ denote electron charge, Planck's constant, and speed of light, respectively) and obtain $EQE \sim 1.5\,\%$ (dashed line in Figure 3c).

$V_{OC}$ scales with $\ln(P_{opt})$, as expected from conventional p-n junction theory. It is related to the MoS$_2$ electron and WSe$_2$ hole carrier densities, $n$ and $p$, by[46] $qV_{OC} = E_g + k_B T \ln[np/N_c^2]$, where $E_g = |E_{CM} - E_{VW}|$ is the effective band gap, $N_c$ is the effective density of states, and $k_B$ denotes Boltzmann's constant. Under steady-state and open-circuit conditions, the carrier concentrations are determined by the balance between photogeneration ($G \propto P_{opt}$) and interlayer recombination ($R$) rates. The recombination rate can, in general, be written by the empirical equation[47] $R = \gamma n^\beta$, where $\gamma$ is a prefactor that is determined by the actual recombination process, $\beta$ denotes the recombination order, and $n = p$. $\beta = 1$ for monomolecular (Shockley Read Hall; SRH) recombination and $\beta = 2$ for bimolecular (Langevin) recombination. By equating $R$ and $G$, we obtain the expression

$$\frac{dV_{OC}}{d\ln(P_{opt})} = \frac{2}{\beta}\frac{k_B T}{q}$$

that we fit to the experimental data (dashed line in Figure 3d) and extract $\beta \sim 1.3$. We conclude that the interlayer recombination is dominated by the SRH process, again evidencing the existence of a large density of trap states in TMDCs.

The power conversion efficiency is defined as $\eta = P_{el,m}/P_{opt}$, where $P_{el,m}$ denotes the output at the maximum power point, and is shown in Figure 3e (blue triangles). The fill factor is calculated according to $FF = P_{el,m}/(J_{SC}V_{OC})$ and is plotted in the same figure (red squares). An efficiency of $\eta \sim 0.2$ % is obtained, which is comparable to values reported for[48] WSe$_2$ (0.1–0.6 %) and[49] MoS$_2$ (1 %) bulk samples, and, more recently, for lateral WSe$_2$ monolayer p-n junctions[9, 10] (0.2–0.5 %). The device can also be operated as a photodiode by biasing it in the third quadrant. At $V = -1$ V, a photoresponsivity of $R = J/P_{opt} = 11$ mA/W is achieved. These numbers need to be judged in light of the weak optical absorption of the ultra-thin 2D structure. A $\sim$10-fold increase may be achieved by stacking several junctions on top of each other, or by plasmonic enhancement[50] of the optical absorption $\alpha$. Moreover, given the $\sim$1.5 % $EQE$ and assuming $\alpha \sim 10$ %, we estimate that $\sim$5 times better efficiency could be obtained by sandwiching the dichalcogenide junction between electrodes for vertical carrier extraction[29, 30].

Our spectrally broad white light source not only excites carriers above the electronic band gap, but part of the spectrum also overlaps with below-gap excitons. These can spontaneously dissociate into free (spatially separated) carriers if the respective band offset is greater than the exciton binding energy, $\Delta E_V$ ($\Delta E_C$) $> E_b$. In order to verify whether this condition is fulfilled, we performed $J_{SC}$ measurements under resonant excitation of the WSe$_2$ ground state exciton at 750 nm wavelength and compared the results with those of above-band gap excitation at 532 nm (see Figure 4a). For a quantitative comparison, we account for the optical interference in the SiO$_2$ layer by plotting the photocurrent against $\kappa P_{opt}$, where $P_{opt}$ is the incident power, as usual, and $\kappa$ accounts for the wavelength-dependent enhancement of the optical intensity (see Figure S9). As depicted in Figure 4a, the photocurrents are similar at typical solar intensities (~1 kW/m$^2 \simeq$ 1 nW), confirming efficient exciton dissociation. The two processes possess different saturation levels, though. We are currently investigating this behavior in more detail.

In Figure 4b we show $J_{SC}$ (red symbols) and $V_{OC}$ (blue symbols) as a function of gate voltage (see corresponding J–V curves in Figure S10). The photocurrent peaks in the p-n regime ($V_G$ = -50 V), whereas it is suppressed for other gate voltages. The reason for this behavior is twofold. (i) A large gate-induced excess carrier concentration (electrons in MoS$_2$ for $V_G \gg$ -50 V; holes in WSe$_2$ for $V_G \ll$ -50 V) reduces the carrier lifetime and gives rise to enhanced recombination of the photo-generated carriers. The drop in open-circuit voltage to both sides of the p-n regime provides evidence for this claim. (ii) The carrier collection efficiency depends on $V_G$ because of the gate-dependent band alignment along the channel of the device.

In summary, given the rapid advances in large-scale production of 2D crystals[5–7], we see potential for using van der Waals heterojunctions in photovoltaic energy conversion. Bringing together different 2D materials in a roll-to-roll process or direct heterostructure growth could lead to a new photovoltaic solar technology. Moreover, due to the plurality of 2D materials with different band gaps and electron affinities, low-cost multi-junction solar cells could come within reach.

*Note added:* During the finalization of this manuscript we became aware of three related studies[51–53].

## ASSOCIATED CONTENT

**Supporting Information**

Raman spectra, scanning electron microscopy image, further electrical characterization, optical excitation spectrum, photocurrent map, optical absorption enhancement, gate-dependent $J$–$V$ characteristics. This material is available free of charge via the Internet at http://pubs.acs.org.

## AUTHOR INFORMATION

**Corresponding Author**

*E-mail: thomas.mueller@tuwien.ac.at. Phone: 43-1-58801-38739.

**Notes**

The authors declare no competing financial interest.

## ACKNOWLEDGEMENTS

We would like to thank G. Kresse, K. Hummer, and K. Unterrainer for valuable discussions, M. Schinnerl, G. D. Cole, and M. Glaser for technical assistance, and E. Bertagnolli and A. Lugstein for providing access to a Raman spectrometer. The research leading to these results has received funding from the Austrian Science Fund FWF (START Y-539) and the European Union Seventh Framework Programme (grant agreement no. 604391 Graphene Flagship).

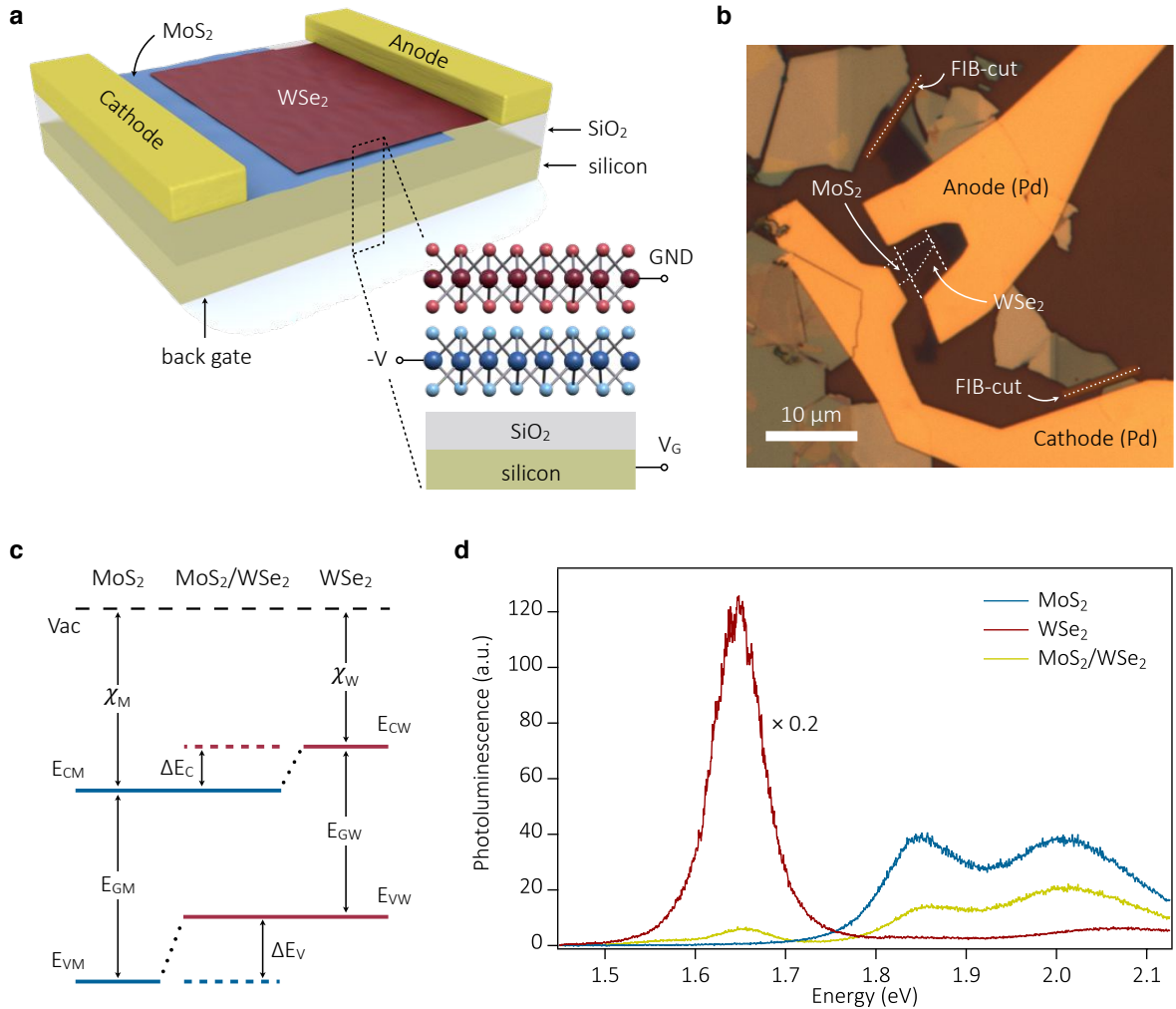

**Figure 1.** (a) Schematic drawing of the device structure. (b) Microscope image of a typical device. Contact electrodes are made of Pd/Au. The $MoS_2$ and $WSe_2$ monolayer flakes are highlighted by dotted and dashed lines, respectively. Short circuits are removed using a focused ion beam (dotted lines). (c) Schematic energy band diagrams of $MoS_2$ (left; blue lines) and $WSe_2$ (right; red lines) in the vicinity of the K-point. The $MoS_2/WSe_2$ heterostructure bands (center) are a superposition of the monolayer bands. The lowest energy electron states (solid blue line) are spatially located in the $MoS_2$ layer and the highest energy hole states (solid red line) lie in the $WSe_2$. The excited states are shown as dashed lines. $E_C$, conduction band edge; $E_V$, valence band edge; $Vac$, vacuum level. (d) Photoluminescence measured at different positions on the sample: $WSe_2$ (red), $MoS_2$ (blue), $MoS_2/WSe_2$ heterojunction (yellow). The photoluminescence from the junction is strongly quenched, due to spatial electron-hole separation/exciton dissociation.

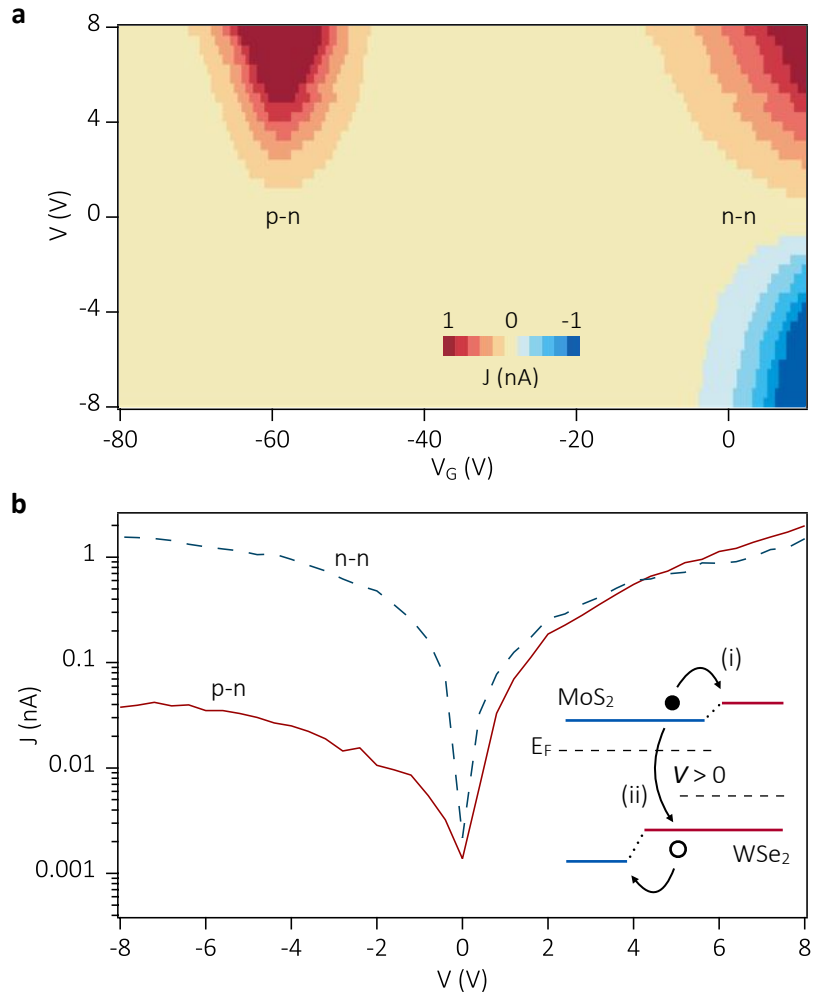

**Figure 2.** (a) Current map, recorded by scanning gate voltage $V_G$ and bias voltage $V$. At $V_G > -10$ V, both flakes are n-type and the device shows resistive behavior. In the range $-71$ V $< V_G < -47$ V, the electrical characteristic exhibits a p-n diode-like behavior. (b) $J$–$V$ traces obtained under n-n ($V_G = +10$ V; dashed blue line) and p-n ($V_G = -59$ V; solid red line) operation (absolute value of the device current $J$ on a logarithmic scale). Inset: Band diagram under p-n operation and forward bias ($V > 0$). Electrons and holes are injected into the $MoS_2$ and $WSe_2$ sheets, respectively, and drift towards the junction where they may (i) overcome the band offsets, or (ii) recombine. $E_F$, Fermi level.

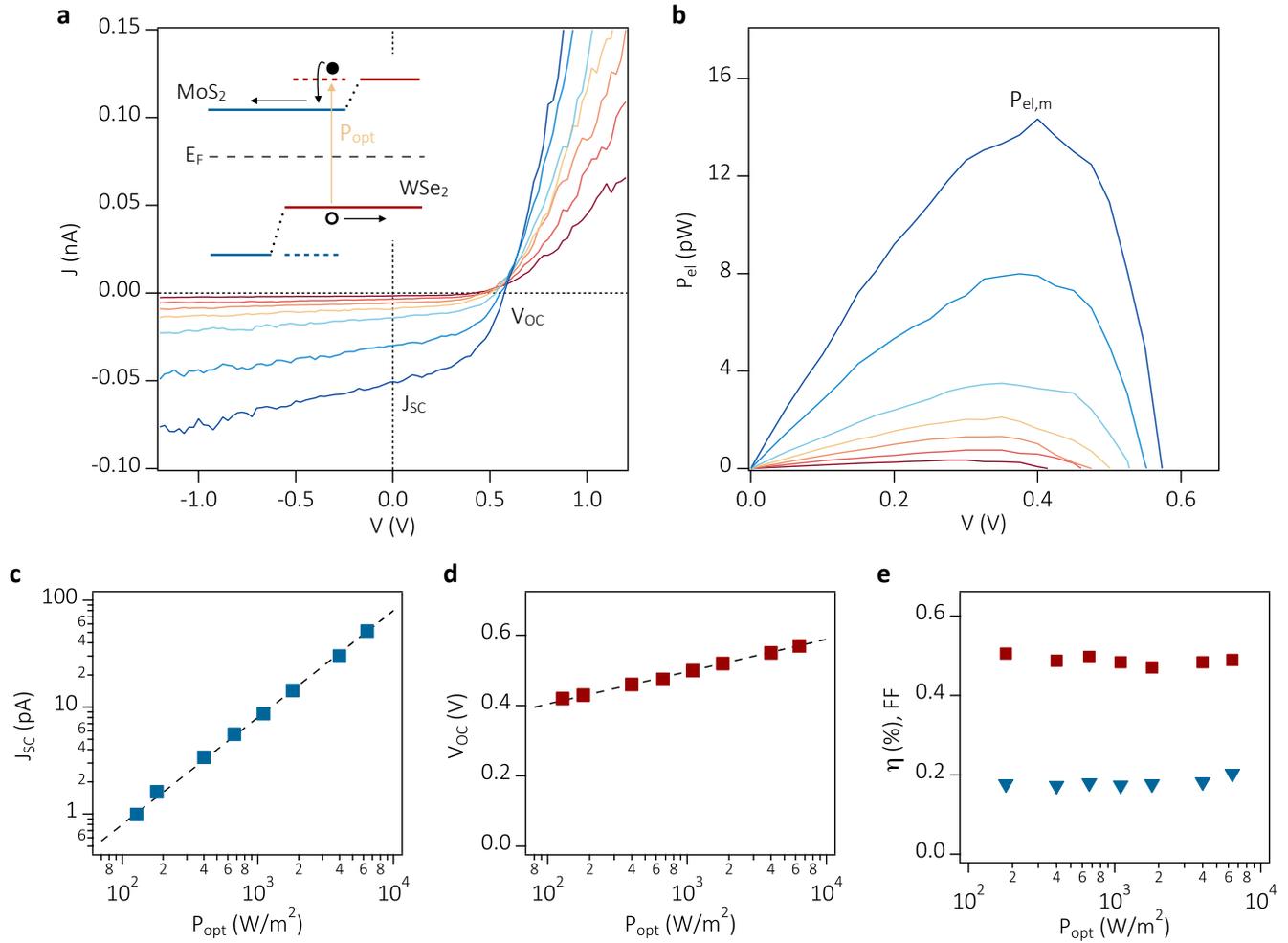

**Figure 3.** (a) *J–V* characteristics of the device under optical illumination with $P_{opt}$ = 180, 400, 670, 1100, 1800, 4000, and 6400 W/m². Inset: Schematic illustration of the photovoltaic effect: electron-hole pairs are generated in the heterostructure, relax to the bottom of the conduction and valence bands, and eventually diffuse to the contacts. (b) Electrical power, $P_{el} = J \cdot V$, that is extracted from the device. $P_{el,m}$ denotes the output in the maximum power point. (c) Short-circuit current $J_{SC}$ and (d) open-circuit voltage $V_{OC}$, as extracted from Figure 3a (symbols, experimental data; dashed lines, theoretical fits). (e) Fill factor *FF* (red rectangles) and power conversion efficiency $\eta$ (blue triangles). During all measurements the gate bias was fixed at $V_G$ = -50 V.

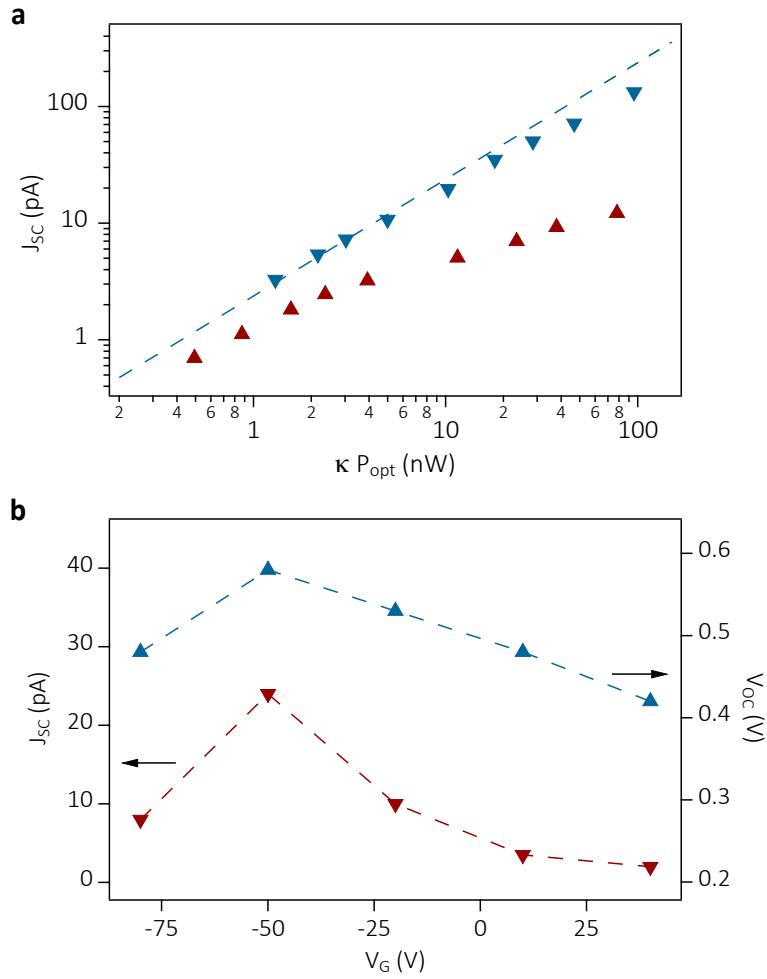

**Figure 4.** (a) Power dependence of short-circuit current $J_{SC}$ for optical excitation at 532 nm (blue symbols) and 750 nm (red symbols). The optical intensity enhancement $\kappa$ (see Figure S9) accounts for standing optical waves in the Si/SiO$_2$ substrate. The dashed line is a fit of $J_{SC} \propto P_{opt}$. (b) Short-circuit current $J_{SC}$ and the open-circuit voltage $V_{OC}$ are shown as red and blue symbols, respectively (white light illumination, $P_{opt}$ = 3 kW/m$^2$).

*Supporting Information*

# Photovoltaic effect in an electrically tunable van der Waals heterojunction

Marco M. Furchi, Andreas Pospischil, Florian Libisch, Joachim Burgdörfer, and Thomas Mueller

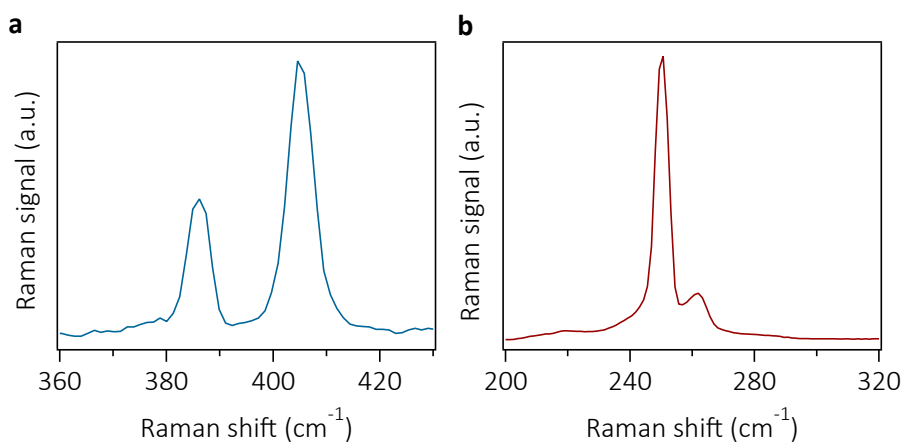

***Figure S1.*** (a) Raman spectrum of $MoS_2$. The energetic splitting between the modes is 19 cm$^{-1}$, indicating monolayer thickness. (b) Raman spectrum of $WSe_2$. The pronounced peak at ~250 cm$^{-1}$, the small shoulder at ~260 cm$^{-1}$, and a missing peak at ~310 cm$^{-1}$ indicate that the sheet is a monolayer.

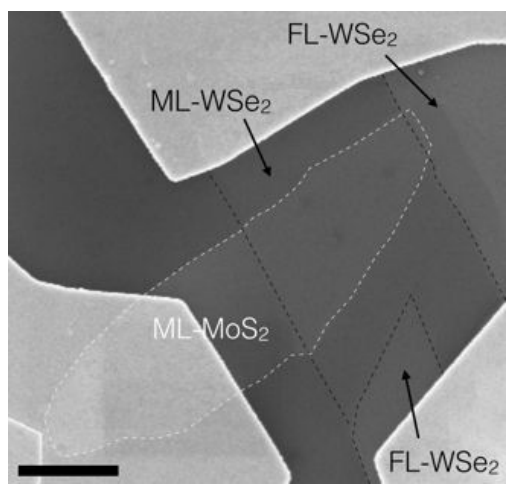

***Figure S2.*** Scanning electron microscopy (SEM) image of Sample I. ML, monolayer flake; FL, few-layer flake. Scale bar, 2 μm.

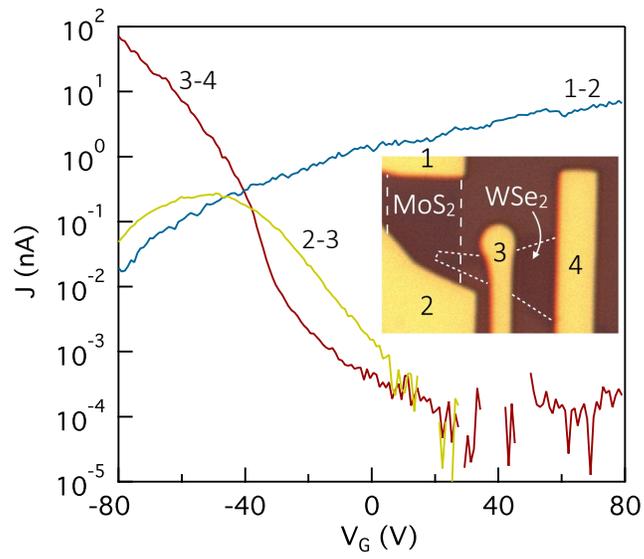

***Figure S3.*** *J-V* characteristics ($V = 3$ V) for the individual MoS$_2$ (measured between contacts 1 and 2) and WSe$_2$ monolayers (contacts 3 and 4), and the heterojunction (contacts 2 and 3) of the device (Sample III) shown in the inset. The formation of a p-n junction at $V_G \sim -50$ V can be understood as a superposition of the MoS$_2$ n-type and WSe$_2$ p-type curves (n-type behaviour of WSe$_2$ ($V_G \gg$) is not observed in this device).

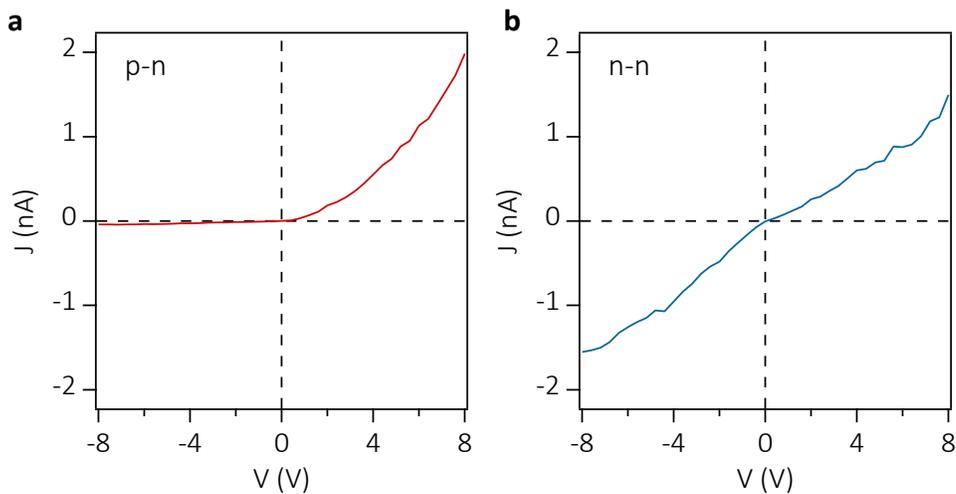

***Figure S4.*** *J–V* characteristics under (a) p-n and (b) n-n configurations (Sample I). The linear behaviour in the n-n regime suggests reliable Ohmic contacts.

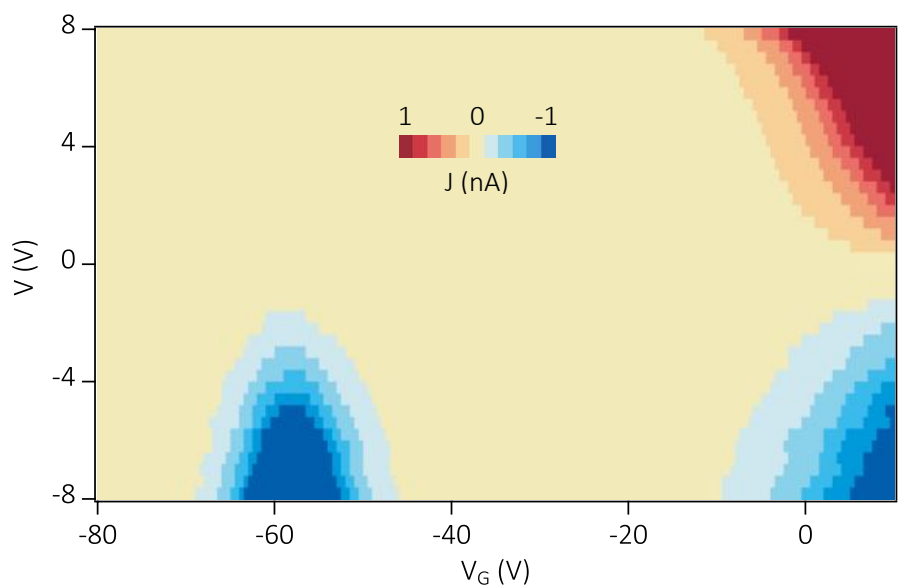

*Figure S5.* Same measurement as in Figure 2a in the paper (Sample I), however, with different electrical wiring: MoS$_2$ sheet grounded; -$V$ applied to WSe$_2$. Voltage and current polarities are reversed, otherwise the device shows same behaviour: the WSe$_2$ sheet acts as anode and the MoS$_2$ as cathode.

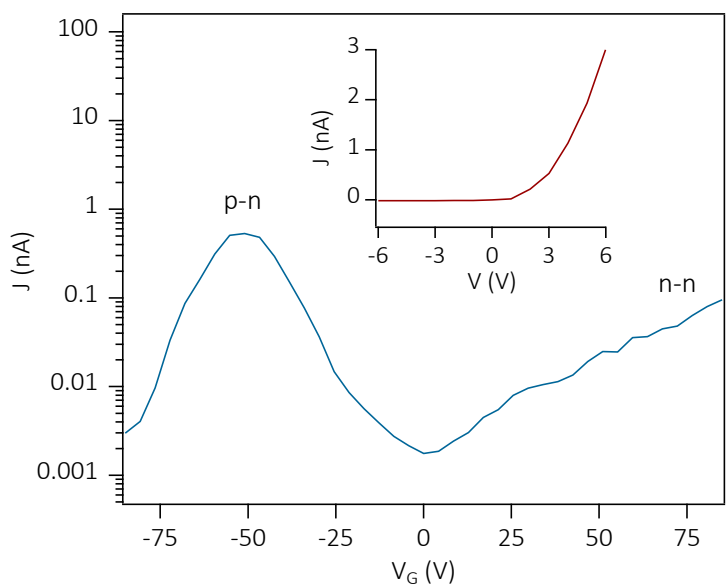

*Figure S6.* Electrical gate-characteristic of Sample II ($V$ = +3 V). Similarly to Sample I, rectifying p-n behaviour is observed at $V_G$ ~ -50 V (see inset); the n-n regime occurs at $V_G$ > 0 V, but is less pronounced.

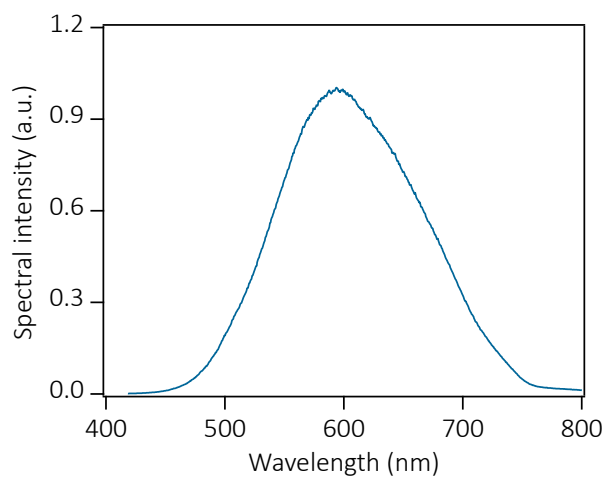

*Figure S7.* Excitation spectrum of halogen lamp.

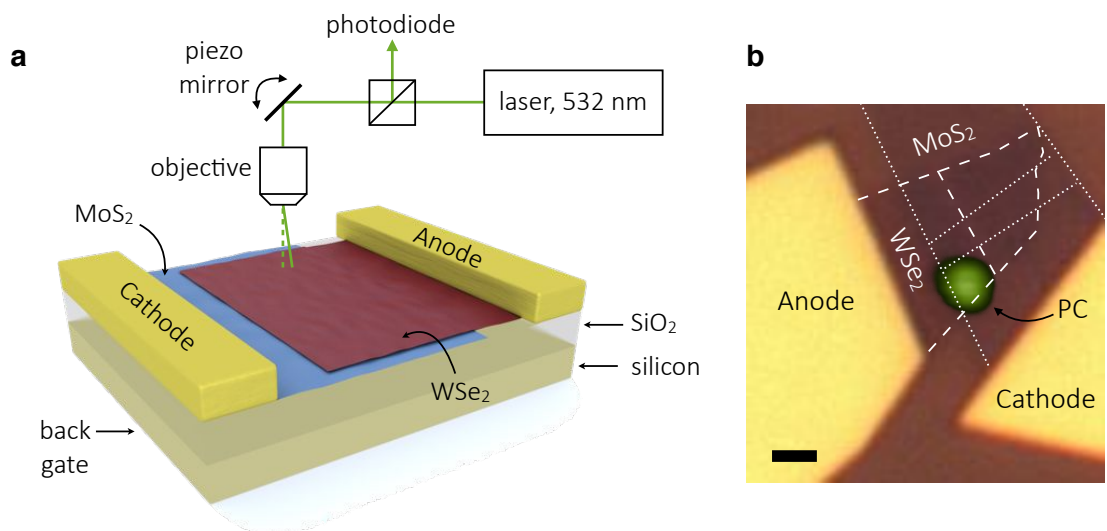

*Figure S8.* (a) A solid-state laser (wavelength: 532 nm) was used as the excitation light source in the photocurrent imaging measurement. The laser beam was focused to a diffraction-limited (~0.6 µm) spot on the device using a microscope objective. A piezo-electrically driven mirror, mounted before the objective, scanned the beam across the sample. The edges of the contact electrodes were first located by detecting the reflected light from the sample with an external photodiode. The incident optical power was adjusted with neutral density filters. (b) Photocurrent image overlaid with an optical micrograph (Sample II). The $MoS_2$ and $WSe_2$ flakes are highlighted by dotted and dashed lines, respectively. The photocurrent (PC) is shown in green and stems from the overlap region between the two flakes. No PC signal is observed in the vicinity of the contacts, where (weak) lateral built-in fields due to Schottky contacts may exist. The flakes micro-crack due to handling, flexing, etc., during exfoliation and/or transfer (also shown as dashed and dotted lines), which currently restricts the area $A$ from which we can collect the photocurrent ($A \sim 1.1$ μm$^2$). Scale bar, 1 μm.

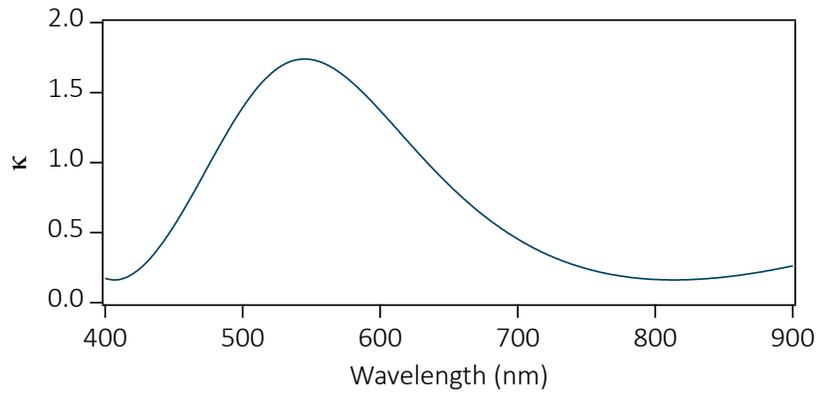

*Figure S9.* Calculated wavelength-dependence of the optical intensity enhancement $\kappa$ due to interference in the 280 nm thick silicon dioxide layer.

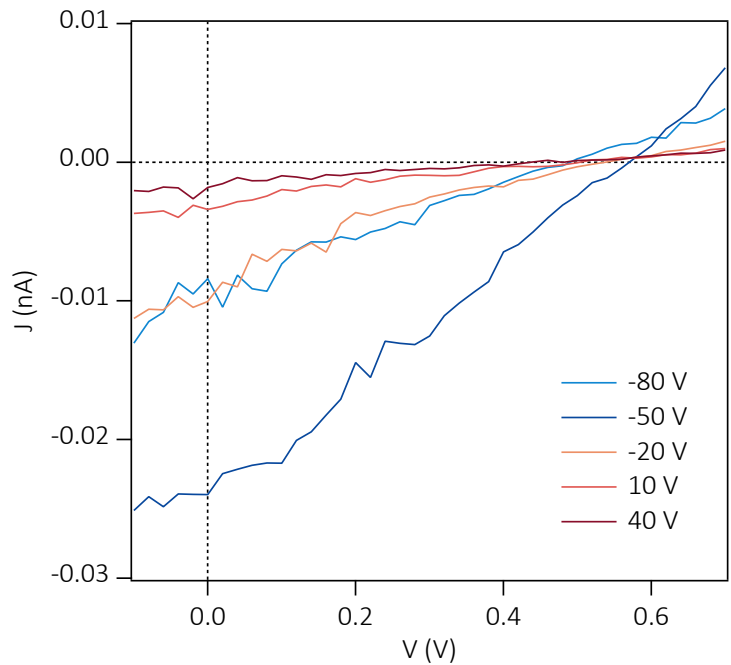

*Figure S10.* J–V characteristics of Sample II under white light illumination with $P_{opt} = 3$ kW/m$^2$ for gate voltages of -80, -50, -20, 10, and 40 V.